# Filamentation-Assisted Isolated Attosecond Pulse Generation

Yu-En Chien[1†], Marina Fernández-Galán[2,3†], Ming-Shian Tsai[1], An-Yuan Liang[1], Enrique Conejero-Jarque[2,3], Javier Serrano[2,3], Julio San Román[2,3], Carlos Hernández-García[2,3], Ming-Chang Chen[1,4,5*]

1. Institute of Photonics and Technologies, National Tsing Hua University, Hsinchu 30013, Taiwan
2. Grupo de Investigación en Aplicaciones del Láser y Fotónica, Departamento de Física Aplicada, Universidad de Salamanca, Salamanca 37008, Spain
3. Unidad de Excelencia en Luz y Materia Estructuradas (LUMES), Universidad de Salamanca, Salamanca 37008, Spain
4. Department of Physics, National Tsing Hua University, Hsinchu 30013, Taiwan
5. National Synchrotron Radiation Research Center, Hsinchu 30013, Taiwan

†These authors contributed equally
* Corresponding author: mingchang@mx.nthu.edu.tw

**Abstract:** Isolated attosecond pulses (IAPs) generated by few-cycle femtosecond lasers are essential for capturing ultrafast dynamics in atoms, molecules, and solids. Nonetheless, the advancement of attosecond science critically depends on achieving stable, high-temporal-contrast IAPs. Our study reveals a universal scenario in which self-compression of the infrared driver in high harmonic generation in extended gas media leads to high-contrast high-frequency IAP generation. Our experimental and theoretical results reveal that filamentation in a semi-infinite gas cell not only shapes the infrared driving pulse spatially and temporally, but also creates a stable propagation region where high harmonic generation is phase-matched, leading to the production of bright IAPs. In an argon-filled gas cell, filamentation notably reduces the pulse duration of Yb-based 1030 nm pulses from 4.7 fs to 3.5 fs, while simultaneously generating high-contrast 200-attosecond IAPs at 70 eV. We demonstrate the universality of filamentation-assisted IAP generation, showing that post-compressed Yb-based laser filaments in neon and helium yield even shorter IAPs: 69-attoseconds at 100 eV, and 65-attoseconds IAPs at 135 eV, respectively. This spatiotemporal reshaping of few-cycle pulses through filamentation possesses immediate impacts on both post-compression techniques and attosecond-based technologies.

The generation of IAPs has revolutionized temporal resolution in light-matter interactions, enabling the observation of the fastest processes such as electron, exciton, and spin dynamics driven by few-cycle pulses[1-4]. Recent studies have further demonstrated that ultrafast lightwaves can induce rapid phase transitions in materials[5,6]. These studies highlight the profound impact of attosecond science, enabling exploration of new physics, uncovering novel mechanisms, and advancing the development of next-generation petahertz devices[7,8].

The development of intense, ultrafast laser pulses, carefully tailored in their spatial and temporal domains, has become a key technology for producing attosecond light pulses via high-order harmonic generation (HHG). This highly nonlinear process is triggered by an intense femtosecond laser pulse interacting with a gaseous or solid target—most typically a gas jet or gas cell. HHG can be understood through the semiclassical three-step model[9,10], which repeats every half-cycle of the driving pulse. First, when the pulse intensity is sufficiently high, an atom is tunnel-ionized. Next, the ejected electronic wavepacket is accelerated in the continuum. Finally, due to the oscillatory nature of the driving field, the wavepacket recombines with the parent ion, emitting high-frequency radiation. With multi-cycle driving pulses, an attosecond pulse train is generated in the time domain, producing multiple harmonic orders in the spectrum. This process occurs in many atoms across the whole target, so harmonic phase-matching needs to be taken into account[11-13]. If the driving pulse is sufficiently short, a single emission event occurs, resulting in the generation of an IAP[14,15]. Alternatively, other techniques relying on polarization gating[16-18] or phase-matching[19-21] can also isolate an IAP from the attosecond pulse train, at the expense of wasting pulse energy.

Currently, the primary method for generating IAPs involves focusing few-cycle carrier-envelope phase (CEP) stabilized pulses—such as 5 fs pulses from a Ti:Sapphire laser (~2 cycles at 800 nm)[22,23] or 12 fs pulses from an optical parametric chirped-pulse amplifier (OPCPA) (~2 cycles at 1800 nm)[24-27], into a short gas cell (SGC). However, the use of a semi-infinite gas cell (SIGC) offers an alternative geometry for HHG[28-30] where nonlinear propagation effects of the driving pulse become significant. When the laser peak power ($P_{peak}$) approaches or exceeds the critical power, intense pulses focused within the SIGC can form a filament, resulting from the balance between Kerr self-focusing, diffraction and plasma defocusing, reshaping the driving electric field both spatially and temporally[31]. Spatially, self-guidance confines the intense field into a narrow channel, overcoming the typical diffraction length obtained in a linear focal geometry. Temporally, the pulse undergoes self-compression. Theoretical calculations suggest that filamentation over an extended propagation distance within the SIGC could enable IAP generation and enhance the high harmonic signal[32-34]. This approach is particularly appealing for its simple setup, requiring only a strong pulse focused into a gas cell to produce IAPs. However, the particular parameters to properly activate the complex spatio-temporal reshaping of the IR pulse during a standard filamentation regime[35,36] explains why, to date, no successful IAP generation driven by filamentation has been reported in the literature.

In this work, for the first time, we experimentally demonstrate robust IAP generation via filamentation driven by Yb-based laser pulses. By focusing a few-cycle driving pulse (< 5 fs at 1030 nm)[37] into argon (Ar), neon (Ne) and helium (He)-filled SIGCs, we show that the resulting HHG supercontinuum exhibits high temporal IAP contrast. Through theoretical simulations, we attribute this result to the self-compression and self-guidance of the driving field induced during the filamentation process in SIGC. Specifically, when 4.7 fs driving pulses are focused into an Ar-filled SIGC, nonlinear reshaping results in a 25% temporal pulse self-compression (from 4.7 fs to 3.5 fs), and in a cleaner spatial mode, leading to high spatial-spectral homogeneity. Compared to conventional SGCs, we demonstrate that IAPs generated in the SIGC exhibit higher temporal contrast and are significantly brighter, indicating that filamentation not only extends the effective propagation distance, but also ensures excellent phase-matching conditions for HHG. Our attosecond streaking measurements further confirm the stability of the filament generated in Ar, demonstrating the emission of bright 200-as IAPs at 70 eV. Additionally, we demonstrate that this filamentation-assisted IAP generation mechanism is universal, as it applies to other gases: 69-as IAPs at 100 eV were generated in Ne, and 65-as IAPs at 135 eV in He. To the best of our knowledge, this is the first study to achieve IAPs directly driven by a post-compressed Yb laser. While Yb-based lasers are known for their robust, high-power femtosecond capabilities, they typically have longer pulse durations (>150 fs), requiring significant post-compression (>40×) to reach few-cycle pulses. This often introduces satellite pulses and wavelength-dependent wavefront distortions that compromise the spatiotemporal pulse quality. Our filamentation-assisted method allows for straightforward spatio-temporal cleaning of these post-compressed pulses into the one-cycle regime without requiring additional dispersion compensation. This unique and reliable approach is highly effective for generating high-contrast IAPs, establishing it as a promising technique for advancing attosecond technology.

**Results**
In our experiment we employed Yb-based CEP-stabilized pulses, with a central wavelength of 1030 nm, post-compressed to a pulse duration of 4.7 fs, and a maximum energy of 500 μJ, operating at a repetition rate of 4 kHz[37]. These pulses were focused to drive HHG into either a SGC or a SIGC using a concave mirror with a 35-cm focal length (see Fig. 1a and Extended Data Fig. 1). On the one hand, the SGC is composed of a stainless steel tube with an inner diameter of 2.2 mm, sealed at one end (Fig. 1b). The long tube of the SGC was aligned perpendicularly to the laser propagation direction and placed near the focus, where the laser itself self-drilled holes through the tube, as shown in Fig. 1b. On the other hand, the SIGC (Fig. 1c) was composed of a gas-filled chamber, with a 0.3-mm thick aluminium plate blocking the interface between the gas-filled and vacuum sections. The focus was positioned near the plate where a hole was self-drilled by the driver beam, allowing the generated high-order harmonics to exit. Ar, Ne and He gases were used to produce high-order harmonics. The resulting few-cycle IR pulses and high-order harmonics were refocused onto a Ne gas jet using an elliptical mirror for attosecond streaking experiments[38] (details are provided in the Methods). Subsequently, an extreme-ultraviolet (EUV) spectrometer was employed to analyze the HHG spectrum.

We first present our experimental results using Ar-filled SGC and SIGC configurations. To prevent overionization, which would reduce the HHG yield due to phase mismatch[12,13,39], we positioned a partially closed iris along the laser beam path before the entrance window. To optimize the high-harmonic supercontinuum flux, we iteratively adjusted the iris aperture, the focal position near the SGC (or near the truncated plate in SIGC), the gas pressure, and the insertion of a wedge pair to control spectral dispersion of the driving pulse. Once the high-harmonic supercontinuum flux was optimized, we obtained the CEP-dependent high harmonic spectra (normalized to their respective global peaks) for both the SGC (Fig. 1d) and SIGC (Fig. 1e). These spectra exhibit a $\pi$ radian periodicity due to the presence of two electric field peaks within one optical cycle[22]. Figure 1f compares the maximum (solid lines) and minimum (dashed lines) CEP-dependent high-harmonic yield in both geometries—green for the SGC, red for the SIGC. In the SGC, the ratio between the maximum and minimum high-harmonic yield is around 2, with the high-harmonic spectra showing noticeable spectral modulations, indicating the presence of multiple attosecond pulses over time. By contrast, in the SIGC configuration the high-harmonic spectrum remains as a clean supercontinuum regardless of CEP variation, while varying in flux. Moreover, the ratio between the maximum and minimum high-harmonic yield in the SIGC exceeds a factor of 10. This supercontinuum and high ratio of CEP-dependent high-harmonic yield directly indicates the generation of high temporal contrast IAPs, as further corroborated through attosecond streaking measurements and theoretical simulations. Notably, the optimized high-harmonic yield in the SIGC configuration is more than twice as bright as that achieved in the SGC, demonstrating the superior performance of the SIGC approach. As a consequence of these results, we shall concentrate here on exploiting HHG in SIGC for bright IAP generation, leaving the details of the SGC results to the extended data.

The optimal conditions for achieving the brightest EUV continuum in the SIGC included an Ar pressure of 150 torr and a partially closed iris, which reduced the input pulse energy from 500 μJ to 203 μJ, resulting in an input $P_{peak}$ of 43.4 GW. Figure 1c illustrates a nonlinear filament extending roughly 6 mm before being truncated by the thin metal plate. Notably, after optimizing the iris aperture and backing pressure for HHG, the $P_{peak}$ of the input pulse approached the critical power $P_{cr}$ of self-trapping, which is given by[40]

$$P_{cr} = 1.8962 \lambda^2 / (4 \pi n_0 n_2), \quad (1)$$

where $n_0$ is the linear refractive index at the central wavelength $\lambda$ and $n_2$ is the nonlinear refractive index of the gas [41], which is directly proportional to the gas pressure. $P_{cr}$ was calculated to be ~ 32.6 GW at 150 torr of Ar, as summarized in Table 1. In short, the optimal conditions for high-harmonic supercontinuum generation (i.e., IAPs) align with those required for stable single filament formation.

To better understand how filamentation enhances the contrast of IAPs, we analyzed the driving beam profile exiting the SIGC filled with Ar as an example. A wedge was inserted after the thin truncated plate to partially reflect the beam out of the vacuum chamber, enabling the measurement of beam profiles in both the near-field and far-field. Figure 2 presents a comparison of the beam profiles exiting the SIGC with 0 torr (first row) and 150 torr of Ar (second row). We assessed the far-field spatial-spectral homogeneity of the output beam after approximately 30 cm of free expansion (central panels of Fig. 2a and 2b) by sampling it with a 1-mm-diameter pinhole along the vertical axis, following the homogeneity criterium based on the spectral overlap as defined in[42]. The near field profiles, obtained via relay imaging, are depicted in the right insets. Figure 2a shows that although the near-field beam profile at the cell output in vacuum shows vertical asymmetry—likely due to astigmatism from the concave mirror—the far-field beam maintains high homogeneity (98.8%). The spot size (or beam divergence) remains nearly constant at ~6.1 mm across the entire spectrum, spanning 600 nm to 1200 nm.

When 150 torr of Ar gas is introduced, nonlinear propagation within the filament reduces the spot size (beam divergence) by ~20%, resulting in a more symmetric beam profile in both the near and far fields (Fig. 2b). Spectrally, the filament induces blue-shift broadening, while the beam diameter remains relatively constant at ~5.5 mm after 30 cm of free expansion. On the contrary, for the red-shifted part above 1200 nm, the beam size increases to approximately 8.9 mm. Additionally, the center of the red-shifted part of the beam (~1200 nm) is vertically displaced by about 1 mm compared to the common center of other wavelengths. These two observations suggest that these longer wavelengths were not guided by the inner part of the filament. Further discussion is provided later. Nevertheless, as the red-shifted part constitutes a very small fraction of the total energy (<10%), the overall spatial-spectral homogeneity remains high (93.5%). The near-field beam, shown in the right inset of Fig. 2b, becomes smaller and more round compared to the one obtained when the cell is in vacuum due to the nonlinear cleaning mode process. Notably, at very high pressures, beam splitting into two was observed as a second filament began to form (see Extended Data Fig. 2), which is expected to negatively impact the high-harmonic flux. This observation points out that optimizing the EUV supercontinuum involves iterative adjustments of the input iris aperture, the focal position within the SIGC, the gas pressure, and the insertion of a wedge pair to control spectral dispersion—essentially equivalent to optimizing single filamentation of the IR for IAP generation.

In order to corroborate the nonlinear nature of the spectral broadening and beam reshaping of our driving beam, we performed theoretical calculations comparing the vacuum and SIGC configurations. To do so, we solve the nonlinear driving beam propagation in the complete volume using as an input the measured experimental pulse before the cell. Our simulation method assumes cylindrical symmetry[43], and includes the complete dispersion of the gas, the Kerr effect, self-steepening and shock terms, and photoionization and plasma absorption. As depicted in Figs. 2c-2d, the numerical simulations replicate the general trends observed experimentally when the SIGC is in vacuum or filled with 70 torr of Ar, respectively. When filled with Ar (Fig. 2d) a general ionization-induced blue-shift in the spectrum is observed, with a consistent spatial width across the whole spectrum. The beam size of the red-shifted portion (>1200 nm) is twice that of other wavelengths, aligning well with our experimental results. From a spatial perspective, we observe a distinctive nonlinear readjustment of the spatial profile in both the near and far fields induced by filamentation. Though the spatial cleaning process is minimal in this case, as a perfect spatial Gaussian profile is assumed at the cell entrance in both cases.

To investigate how filamentation reshapes the temporal profile of the driving pulse, we employed the TIPTOE method[44-46] (more details are provided in the Methods), which precisely characterizes the femtosecond pulse waveform. Figures 3a and 3b compare the IR waveforms at the SIGC exit with (blue) and without (red) 150-torr Ar, as obtained from TIPTOE. The negative values on the time axis indicate the leading time. In Figs. 3c and 3d we present the half-cycle-by-half-cycle temporal delay and the corresponding frequency shift between the two waveforms shown in Fig. 3a. Our analysis divides the time axis into five segments: I. Before -30 fs, both waveforms are nearly identical, as the electric field is too weak to induce nonlinear effects. II. Between -30 fs and -20 fs, we observe a red-shift in the waveform when the cell is filled with Ar. We attribute it to the Kerr effect, where the light wave undergoes a nonlinear phase *delay* caused by its own intensity. III. After -20 fs, in the central part of the pulse, the strong electric field ionizes Ar. Since the plasma refractive index is lower than 1, the light wave experiences a phase *advance*, resulting in a spectral blue shift. Specifically, different ionization rates between half-cycles lead to varying blue shifts within each half-cycle, effectively shortening the pulse duration (as shown in Fig. 3b). IV. After 10 fs, as the electric field weakens, the Kerr effect again dominates, causing another spectral red-shift. V. After 20 fs, similar to segment I), the laser intensity is insufficient to cause any changes in the waveform. As a result, the spectrum of the pulse when the SIGC is filled with Ar experiences significant blue-shift with some red-shift, leading to pulse self-compression from 4.7 fs to 3.5 fs, as shown in Fig. 3e (temporal domain) and Fig. 3f (spectral domain). Our theoretical calculations for the nonlinear driving field propagation (Figs. 3g and 3h) show strong agreement with our main experimental findings.

It is worth noting that the red-shifted spectrum resulting from filamentation-based post-compression exhibits unique characteristics. Spatially, the beam divergence around 1200 nm is significantly greater than for wavelengths below 1200 nm (see Figs. 2b and 2d). Spectrally, a distinct phase shift behavior is observed around 1200 nm (see Figs. 3f and 3h). The Fourier analysis presented in Extended Data Fig. 3 also indicates that the red-shifted component primarily originates outside the main pulse—mainly from -20 fs, corresponding to segment II, and partially from +20 fs, corresponding to segment IV in Fig. 3d—where the local peak intensity induces self-phase modulation (SPM) but is insufficient to sustain filamentation. In contrast, the blue-shifted component arises from time segment III (the main pulse), where the peak intensity is sufficient to induce filamentation, enabling spatial guidance and temporal compression. A similar analysis in the SGC configuration is displayed in Extended Data Fig. 4. Notably, Like the SIGC, the SGC exhibits a self-compression effect; however, it lacks a guiding effect, making the IR beam prone to splitting.

In addition to Ar, we utilized Ne and He gases in the SIGC for HHG. After the same optimization procedure in Ne (He) SIGC, a partially closed iris reduced the input energy to 355 μJ (420 μJ), resulting in $P_{peak}$ of approximately 88.8 GW (127.3 GW). The optimal Ne (He) pressure for achieving the brightest high-harmonic supercontinuum was found to be around 650 torr (1600 torr). At these pressures and considering the corresponding nonlinear refractive index $n_2$[32,47], Eq. (1) estimates the critical power $P_{cr}$ to be ~108.3 GW (~162.5 GW). A comparison between their $P_{peak}$ and $P_{cr}$ is summarized in Table 1. Surprisingly, in all three independent SIGC experiments—Ar, Ne, and He—the optimal $P_{peak}$ for generating the high-harmonic supercontinuum approached the theoretical $P_{cr}$ required for stable single filament formation. As expected, similar to the experimental results with Ar, we observed self-guidance of the IR beam in space and self-compression of the IR pulse in time when using Ne and He (Extended Data Figs. 5 and 6), though the degree of compression and guidance was less pronounced than with Ar. We emphasize that when using the 2.2-mm SGC, the brightest HHG from Ne was approximately 13 times weaker than that generated in the SIGC. Furthermore, we were unable to generate HHG from He in the SGC, even at very high gas pressures. These findings clearly demonstrate the critical role of spatiotemporal shaping of the IR beam, induced during the filamentation process, for the successful generation of bright IAPs in the SIGC.

After characterizing the nonlinear filamentation in the SIGC, we present in Fig. 4 the results of EUV attosecond streaking experiments conducted in Ar (first row), Ne (second row) and He (third row). The experimental setup is shown in Fig. 1a. By focusing both IR and EUV pulses into a Ne gas jet, and introducing a time delay between them, the momentum shift of the photoelectron across the entire photoelectron spectrum was recorded (first column in Fig. 4). The presence of only one distinct streaking trace confirms the generation of high-contrast IAPs. The IAPs generated in the SIGC filled with Ar (Ne and He) have central energies of 65 eV (100 eV and 135 eV), bandwidths of ~10 eV (~25 eV and ~35 eV), and can support transform-limited pulse durations of 163 as (58 as and 42.8 as). Using the retrieval algorithm PROOF[48] (details presented in Extended Data Fig. 7), the measured durations of the IAPs generated from Ar (Ne and He) are 203 as (69 as and 65 as). The presence of an IAP in all three experiments is further corroborated by the CEP-scan HHG spectra shown in Extended Data Fig. 8. The streaking experiments confirm that both the filamentation-based post-compressed IR and the IAPs are exceptionally stable. This result also marks the first demonstration of IAPs driven by a post-compressed Yb laser (see Extended Data Fig. 9).

**Discussion**

Our experimental results have demonstrated filamentation-assisted IAP generation using Yb-based few-cycle laser pulses undergoing self-compression and self-guidance in Ar, Ne and He. Unlike in the SGC configuration, the self-cleaning effect created by the filament plays a crucial role for the buildup of the harmonic signal along the SIGC, enabling efficient harmonic phase-matching. To gain a comprehensive understanding of the HHG process, we have performed complete numerical simulations. Our method combines the nonlinear driver propagation and the single-atom HHG calculations with the electromagnetic field propagator, which takes into account the integral solution of the Maxwell equations[49]. At the single-atom microscopic level, we solve the time-dependent Schrödinger equation (TDSE) under the commonly used single-active electron approximation. It is important to note that approximations like the strong field approximation, while faster, fail to accurately reproduce HHG driven by few-cycle laser pulses like those used in our experiment[50]. As commented above, we also account for the nonlinear propagation of the driving field using the method described in[43], as well as for harmonic propagation and phase-matching. Note that this unique combination of nonlinear propagation and HHG has been successfully used in lower-pressure configurations were IAPs were obtained[30].

Figure 5a shows key parameters of the driving field as it propagates nonlinearly through the output end of the SIGC filled with 70 torr of argon. Due to the self-guiding of the driver during the filamentation process, the beam radius (blue) remains relatively constant at the final part of the SIGC. In contrast, the beam in the SGC diverges much more rapidly (similar analysis for SGC shown in Extended Data Fig. 10). We depict the peak intensity (green) and pulse duration (red) for the pulse on-axis. Overall, these results suggest favourable harmonic phase-matching conditions due to the excellent spatio-temporal characteristics of the driver (extremely short and almost perfectly collimated) in this region of the SIGC, comparable to those achieved in gas-filled waveguides[11,51-53] or to the vertical phase-matching branch described in[39]. These favourable conditions are evidenced in Fig. 5b, were HHG build-up is shown through the 3D-TDSE single-atom HHG temporal emission (top) and spectra (bottom) at every on-axis position across the propagation direction in this final region of the SIGC. Nonlinear propagation ensures clean HHG generation, especially near the focus position ($z = 0$). However, to take into account the complete HHG build-up, off-axis contributions and transverse phase-matching also need to be considered besides longitudinal phase-matching.

Figure 5c presents the results of complete macroscopic propagation simulations, incorporating both longitudinal and transverse phase-matching. The temporal (top) and spectral (bottom) results confirm the generation of an IAP as short as 320 as, along with a secondary burst with less than 10% of peak intensity. This secondary burst introduces modulations into the correspondent HHG spectrum, which were not observed in the experimental results. We believe that the experimental optimization procedure consisting on an iterative adjustment of the the input iris aperture, the focal position in the SIGC, the gas pressure, and the insertion of a wedge pair to control spectral dispersion, which is computationally unfeasible to mimic, have helped to achieve the optimum transverse phase-matching conditions, similar to those reported in gas jet experiments[54,55].

In summary, we have demonstrated that filamentation-assisted IAP generation in the SIGC enhances the temporal resolution of few-cycle IR pump and IAP probe experiments, by simultaneously shortening the few-cycle pulse and achieving high-contrast IAPs. We observed that a SIGC configuration more effectively facilitates IAP generation compared to SGC. Moreover, this versatile method is applicable across common inert gases used for HHG. Notably, IAPs generated in this work are driven by a turn-key Yb laser, which have become the mainstream high-power femtosecond sources. Therefore, the nearly maintenance-free IAP source demonstrated here promises to enhance accessibility and advances attosecond science and technologies. The filamentation-based self-compression scheme also shows promise for preserving pulse quality while supporting bandwidths beyond an octave—surpassing the limitations of chirped mirrors. This approach could enable multistage setups, paving the way for sub-femtosecond, high-intensity IR pulses and advancing ultrafast science.

**Table 1 | Optimized peak power for high harmonic supercontinuum vs. theoretical critical power for filamentation.**

| Gas | Optimized Peak Power $P_{peak}$ | | | Theoretical $P_{cr}$ | | | | $P_{peak} / P_{cr}$ |
|---|---|---|---|---|---|---|---|---|
| | Pulse energy (μJ) | Pulse duration (fs) | Peak power (GW) | Center wavelength (nm) | Optimized pressure (torr) | $n_2$ at 1 atm ($\times 10^{-19}$ cm$^2$/W) | Critical power (GW) | |
| Ar | 203 | 4.68 | 43.4 | 924 | 150 | 2.0[41] | 32.6 | 1.33 |
| Ne | 355 | 4.00 | 88.8 | 927 | 650 | 0.14[32] | 108.3 | 0.82 |
| He | 420 | 3.30 | 127.3 | 928 | 1600 | 0.038[47] | 162.5 | 0.78 |

**Figures and Tables**

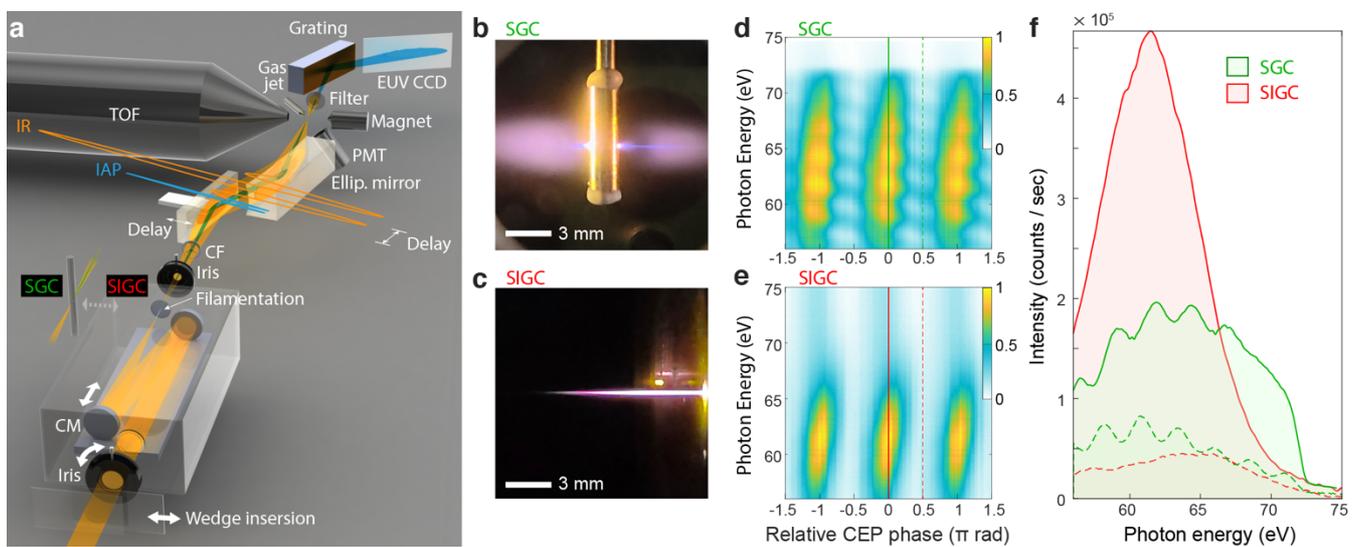

**Fig.1 | Experimental setup and CEP-dependent high harmonic spectra in SGC versus SIGC**. (**a**) Schematic representation of HHG in SGC (**b**) versus SIGC (**c**), and pulse characterization using TIPTOE and attosecond streaking. (**d**) and (**e**) display the normalized CEP-dependent HH spectra, driven by post-compressed < 5 fs pulses from an Yb laser, in SGC and SIGC, respectively. The solid line indicates the optimized CEP (relative CEP phase = 0) for maximum HHG yield. A relative change of 0.5 rad in CEP (dashed lines) results in significant variation in high harmonic yield. (**f**) shows a direct comparison of these CEP-dependent HH yields. Abbreviations: CM - concave mirror; TOF - time of flight; CF - central filter with IR passing on the outer side and a 2-mm diameter thin metal filter (e.g., 200-nm Al, Zr, or Ag) suspended in the center by three thin wires, blocking IR while allowing HHG to pass through.

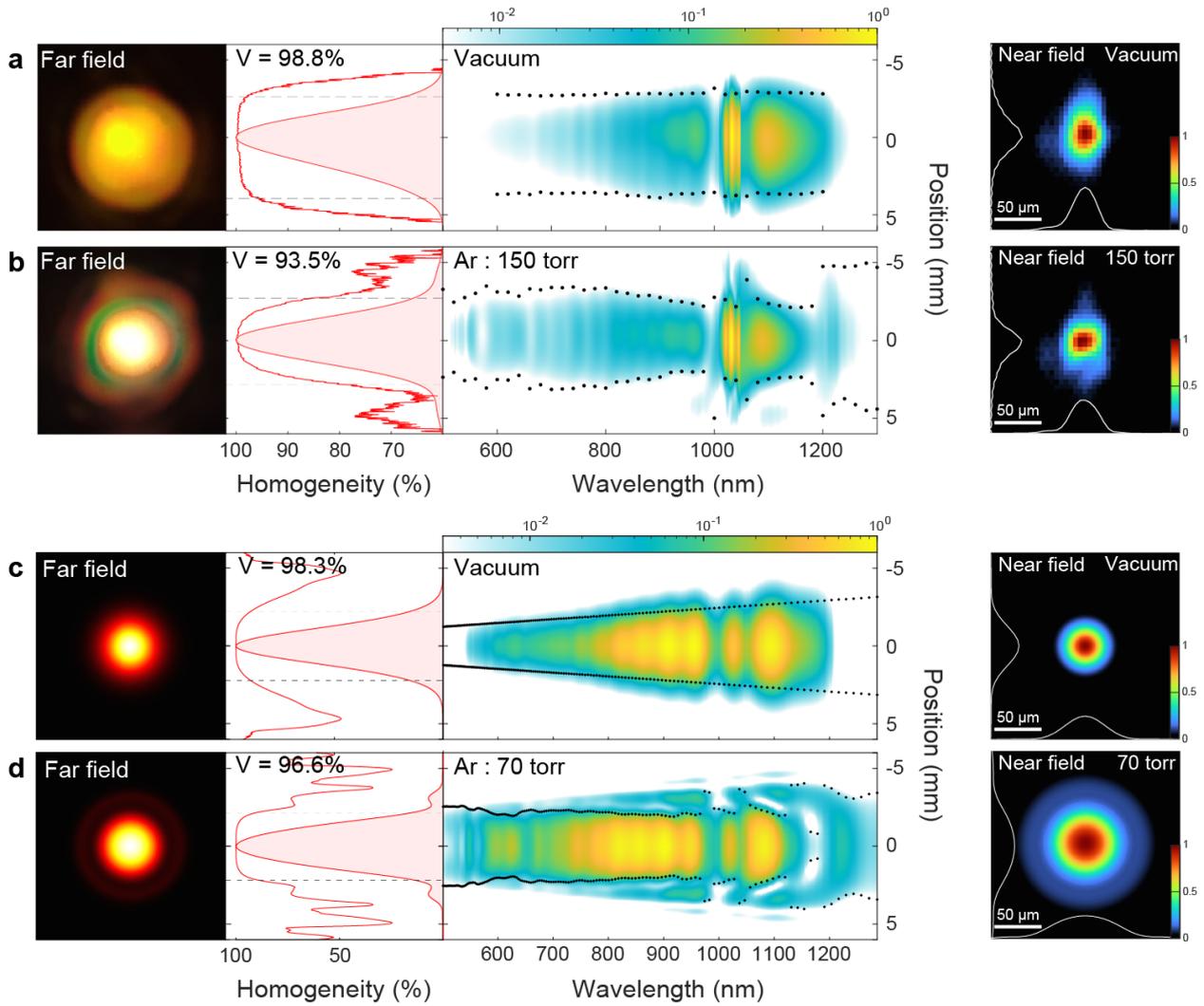

**Fig. 2 | Self-guiding and self-cleaning effects in IR filamentation.** The first column presents a direct comparison of the far-field beam profiles from the SIGC: (**a**) under vacuum and (**b**) with 150 torr of Ar. The second column displays their normalized integrated power and spatial-spectral beam homogeneity. The third column illustrates the spatial-spectral distribution of the output beam along the vertical axis. Dotted lines in the second and third columns indicate the beam waist at $1/e^2$ level. The fourth column depicts the mode profile of the IR beam at the exit of the gas cell, captured via relay imaging using a standard Si-based CMOS sensor. (**c**) and (**d**) show corresponding beam profiles and spatial-spectral distributions from numerical simulations under vacuum and with 70 torr of Ar, respectively.

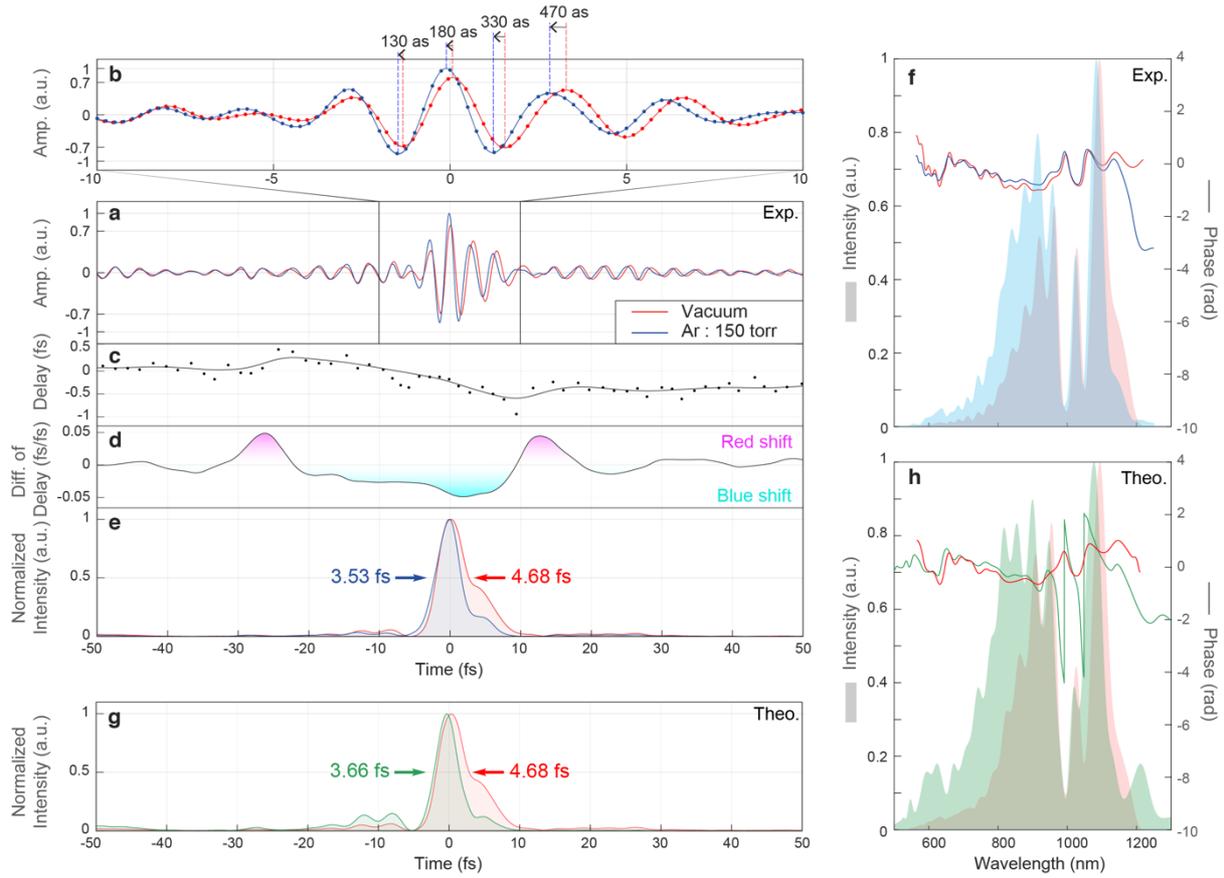

**Fig.3 | Filamentation-induced pulse self-compression.** Waveform measurements were conducted using TIPTOE. (**a**) The waveform of the few-cycle pulse propagating through a SIGC in vacuum (red lines) versus when filled with 150-torr Ar (blue lines), which is the optimized pressure for producing the brightest high-harmonic supercontinuum. To facilitate comparison, the curves are relatively normalized to the time point t = -50 fs. (**b**) Zoomed-in view of **a**, showing the half-cycle-by-half-cycle blue-shift of the waveform induced by ionization. Experimental data points, indicated by dots, have a temporal resolution of 250 as. Notably, the standard deviation across eight experimental trials is minimal, comparable to the size of the dots. (**c**) Half-cycle by half-cycle temporal delay extracted from **a** (dots: raw data; solid line: smooth curve weighted by the field amplitude). (**d**) Frequency shift derived from **c**. (**e**) and (**f**) show the corresponding normalized intensity profile and spectrum, respectively. The blue-shift results in pulse self-compression from 4.7 fs to 3.5 fs. (**g**) and (**h**) show the normalized intensity profile and spectrum obtained from numerical simulations.

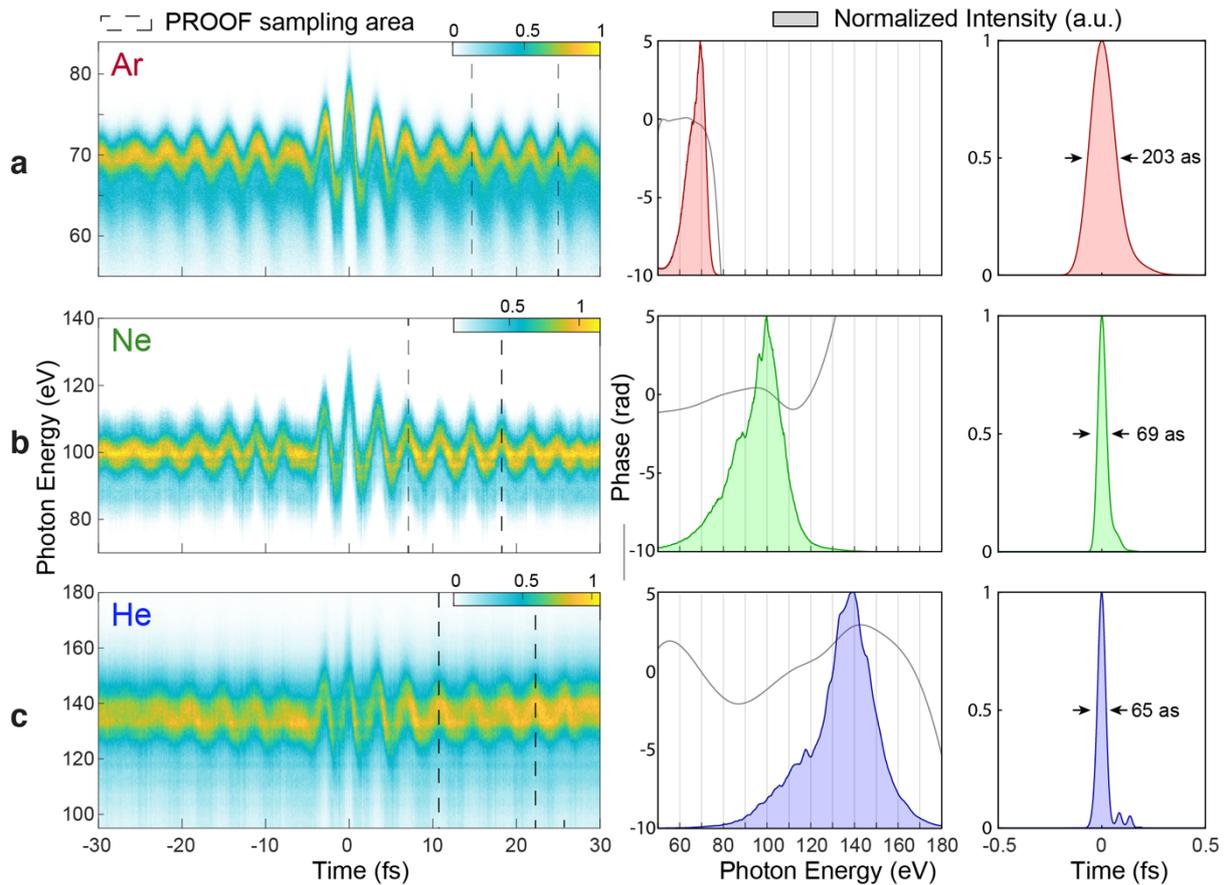

**Fig. 4 | Attosecond streaking results.** Experimental spectrograms were obtained from self-compressed IR pulses and the resulting IAPs generated from filaments in (a) Ar, (b) Ne, and (c) He. The middle panel displays their EUV spectra along with the phase of the resulting IAPs, while the pulse durations are shown in the right panel. The measured durations of the IAPs generated from Ar, Ne, and He are 203 as, 69 as, and 65 as, respectively, compared to their transform-limited pulse durations of 163 as, 58 as, and 42.8 as. All pulses exhibit a slight positive chirp. The PROOF algorithm was employed to retrieve these results[48].

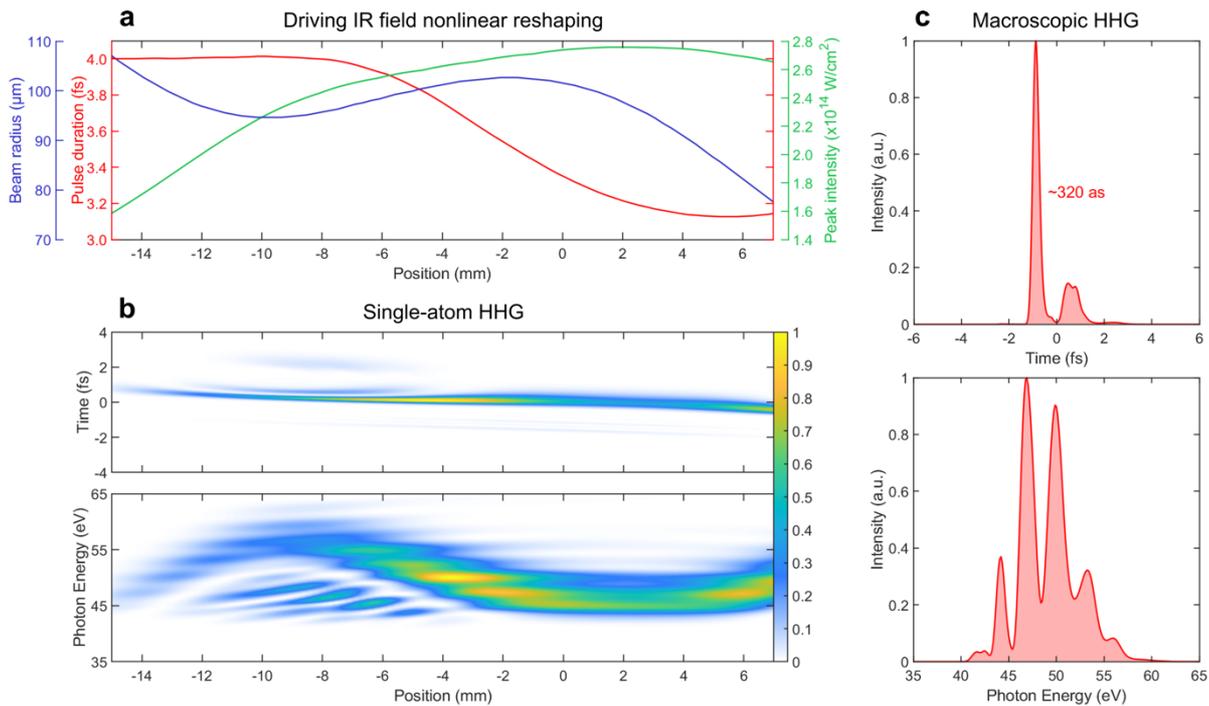

**Fig. 5 | Numerical simulations of nonlinear driving field propagation and IAP generation in a SIGC.**
(**a**) Peak intensity (green) and size (blue) of the driving beam as it propagates along the trailing edge of the SIGC filled with 70-torr Ar. The temporal FWHM of the driving pulse that propagates on-axis is plotted in red. For the simulation we used a beam with 203 µJ input energy with the same temporal profile as measured in the experiment and with a Gaussian spatial shape. With the SIGC evacuated, the beam has a waist of 40 µm at the focus (position z = 0 in the horizontal axis), which is located 7 mm before the end of the SIGC. (**b**) 3D-TDSE single-atom HHG calculations along the SIGC propagation axis reflecting the attosecond temporal emission (top panel) and the corresponding harmonic spectra (bottom panel). (**c**) Attosecond pulse (top) and its EUV spectrum (bottom) obtained from the full macroscopic HHG simulations after propagation to a far-field detector. These macroscopic calculations predict the emission of a prominent isolated attosecond burst with a FWHM duration of ~320 as. A secondary burst appears with <10% intensity compared to the main one, resulting in spectral modulations.

## Methods

**Attosecond streaking:**

An attosecond streak camera, shown in Fig. 1a, was used to analyze the filamentation-based post-compressed IR pulse and the resulting EUV pulses. After the SIGC, the EUV and residual IR passed through a customized filter: IR was transmitted through the outer section, while a 2-mm diameter thin metal filter (e.g., 200-nm Al, Zr, or Ag) suspended at the center by three thin wires blocked the IR, isolating the EUV in the center. The delay between the central EUV and annular IR was adjusted by moving the outer long mirror on a spatial delay stage. It consists of a long reflective mirror (30 mm by 120 mm) with a slot in the middle, into which a smaller long reflective mirror (2 mm by 20 mm) is inserted. This specialized beam splitter, comprising two long mirrors, is coated with Ag and operates at a grazing incidence angle of 6°, providing high reflectivity for the IR beam. Additionally, an iris was incorporated into the setup to control the IR intensity, optimizing the streaking addressing field. Both the EUV and IR beams were then reflected by a Ni-coated elliptical mirror (with a focal length of 20 cm) at a grazing angle of 6° and focused into a Ne gas jet, where the EUV beam generated photoelectrons. These photoelectrons were collected using a 60-cm-long magnetic bottle electron time-of-flight spectrometer, and the recorded spectrogram captured the time delay between the EUV and probe pulses. For precise delay control, a piezoelectric transducer, based on interference fringes from a CW 808 nm laser, was used to feedback control the delay between the EUV and IR pulses. The uncertainty in the relative delay is as low as approximately 20 as.

**TIPTOE:**

The TIPTOE setup[44-46] is similar to the attosecond streaking setup, shown in Fig. 1a, with two modifications: 1) the central IR filter (a 2 mm diameter metal foil, such as Al, Zr, Ag) was removed, and 2) ~30 torr of Ar were introduced into the streaking chamber to fully absorb the generated EUV radiation. A spatial beam splitter was used to divide the IR beam into central and annular parts. The delay between the central and annular beams was adjusted by moving the outer long mirror. After passing through the beam splitter, the two beams were focused by an elliptical mirror. The intense annular beam ionized the ~30 torr of Ar at the focus, while the central beam perturbed the ionization yield. To prevent overionization in the air, a partially closed iris was placed along the annular beam path. The ionization fluorescence was collected by a photomultiplier, and a boxcar integrator was used to improve the signal by averaging 100 ionization events. The modulation of the ionization yield, measured as a function of the time delay between the two pulses, provided a direct representation of the electric field of the pulse. All TIPTOE data presented here were obtained after spectral filtering in the range of 500 nm to 1300 nm. Note that when the chamber is filled with 30 torr of Ar, the distance from the exit of the filamentation to the TIPTOE focus is approximately 120 cm. The corresponding GDD is only 0.8 fs², which hardly stretches the 3.5 fs IR pulse.

**Acknowledgments:** M.-C.C. acknowledges Taiwan's 2030 Young Scholar Fellowship. We gratefully acknowledge funding support by National Science and Technology Council grants 113-2112-M-007-042 and 112-2112-M-007-052 to M.-C. C. We also received funding from Ministerio de Ciencia e Innovación (PID2022-142340NB-I00), from the European Research Council (ERC) under the European Union's Horizon 2020 Research and Innovation Program (Grant Agreement No. 851201), and Junta de Castilla y León and Fondo Europeo de Desarrollo Regional (FEDER), under grant No. SA108P24. M.F.G. acknowledges support from Ministerio de Universidades under Grant FPU21/02916. The authors thankfully acknowledge RES resources provided by BSC in MareNostrum 5 and CESGA in Finisterrae III to FI-2024-2-0010 and FI-2024-3-0035.

**Author Contributions:** Y.-E. C., M.-S. T., and M.-C. C. designed and constructed the experimental setup. Y.-E. C. and M.-S. T. collected the data. Y.-E. C., A.-Y. L., and M.-C. C. analyzed the TIPTOE and streaking data. M. F.-G., E. C.-J., J. S., J. S. R., and C. H.-G. performed simulations of nonlinear propagation and HHG. All authors contributed to writing the manuscript, which was first drafted by Y.-E. C. M. F.-G., J. S. R., C. H.-G. and M.-C. C. All authors discussed and interpreted the experimental data.

**Competing interests**: The authors declare no competing interests.

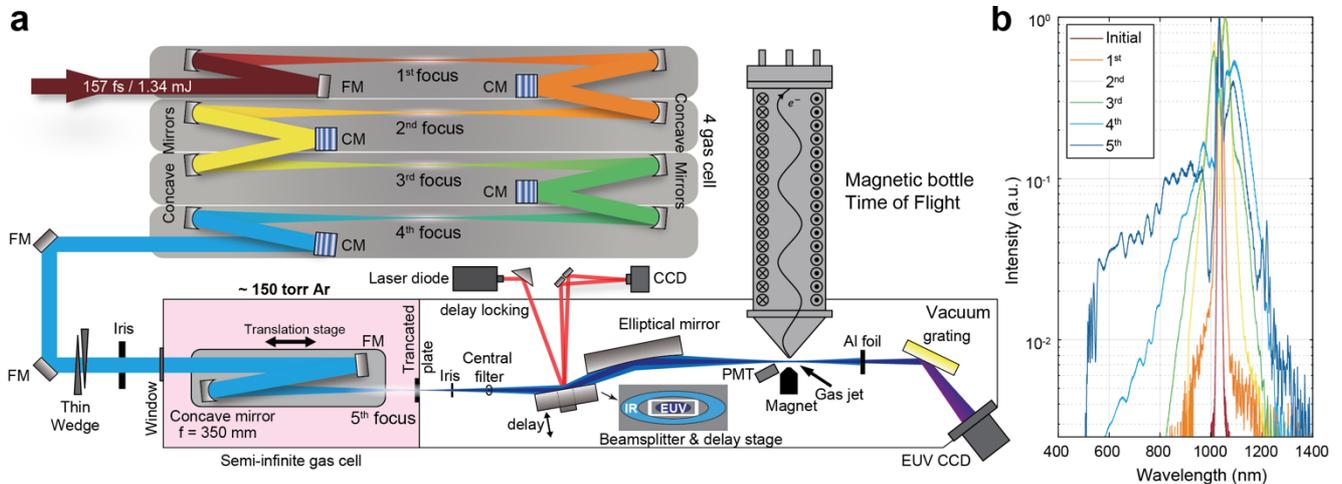

**Extended Data Fig. 1 | High-contrast single-cycle pulse generation through post-compression with five focuses.** (**a**) The first four focuses utilize SPM for spectral broadening, combined with dispersion engineering, to compress the Yb laser pulse from ~160 fs to less than 5 fs. The fifth focus employs filamentation-assisted pulse shortening, where ionization-induced self-compression near the pulse peak further reduces the pulse duration and enhances pulse contrast. Notably, unlike the previous stages, no additional dispersion management is required for this final pulse shortening. This filamentation process simultaneously generates IAPs. For measuring the IAP pulse duration, the beam passes through an iris (which controls the strength of the addressing field in streaking experiments), a central filter (200 nm Al, Zr, or Ag), and a spatial beam splitter, which creates a delay between the EUV (central beam) and IR (annular beam). The delay is monitored using an interferometer, and with PID control, the delay between the EUV and IR pulses is adjustable with an uncertainty of less than 20 attoseconds. The EUV and IR beams are simultaneously focused onto a Ne gas jet using an elliptical mirror, where the photoelectron spectrum is analyzed by a magnetic bottle time-of-flight spectrometer. Following this, an EUV spectrometer, consisting of a concave grating and an EUV CCD, is used for HHG analysis. (**b**) The spectrum of the incident Yb laser and the spectra after passing through each focus. CM: chirped mirror, FM: flat mirror, PMT: photomultiplier tube.

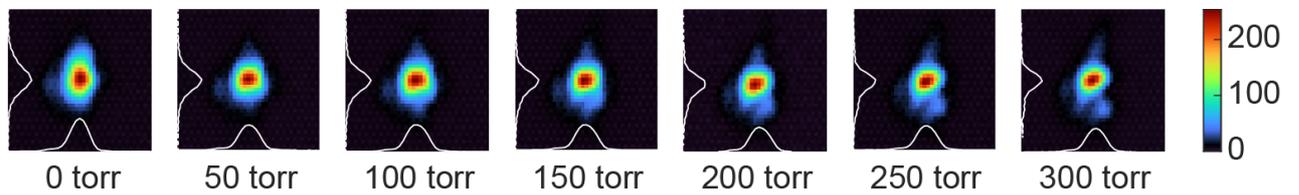

**Extended Data Fig. 2 | Second filament formation in Ar-filled SIGC.** The mode profile of the IR beam at the exit of the gas cell, captured via relay imaging, as the Ar pressure in the SIGC is gradually increased. The optimal pressure for generating the EUV supercontinuum is 150 torr. As the pressure continues to increase beyond this point, the emergence of a second filament becomes visible in the lower right of the beam profile. These images were acquired using a standard Si-based CMOS sensor.

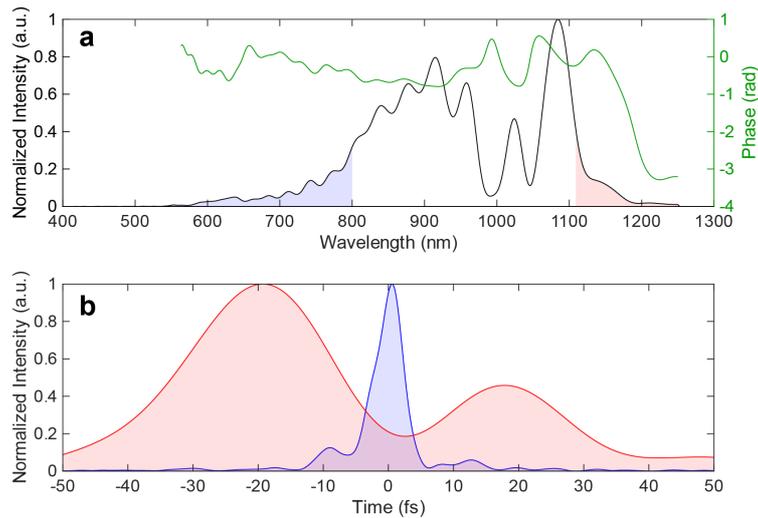

**Extended Data Fig. 3 | Origin of the reddish and bluish spectra from filamentation.** (**a**) The black and green curves represent the spectrum and spectral phase, respectively, from the filamentation-based post-compression in an Ar-filled SIGC (Fig. 3f, obtained via TIPTOE). The red and blue shaded areas indicate the most red- and blue-shifted components of the spectrum produced in this process. The inverse Fourier transform of these spectral regions reveals their temporal positions, as shown in (**b**). The red-shifted component originates from outside the main pulse, while the blue-shifted component aligns with the main pulse (peaking at 0 fs). The blue-shifted component undergoes self-compression, requiring no additional dispersion engineering (e.g., customized chirped mirrors). Although the red-shifted component lies outside the main pulse, it contributes a very small proportion of energy when using few-cycle pulses (< 10% in our case). Another important point to emphasize is that filamentation-based post-compression can achieve self-compression for a bandwidth exceeding one octave, which is beyond the capability of chirped mirrors. Consequently, the traditional approach of combining SPM with negative chirped mirrors becomes ineffective for ultrabroadband spectra.

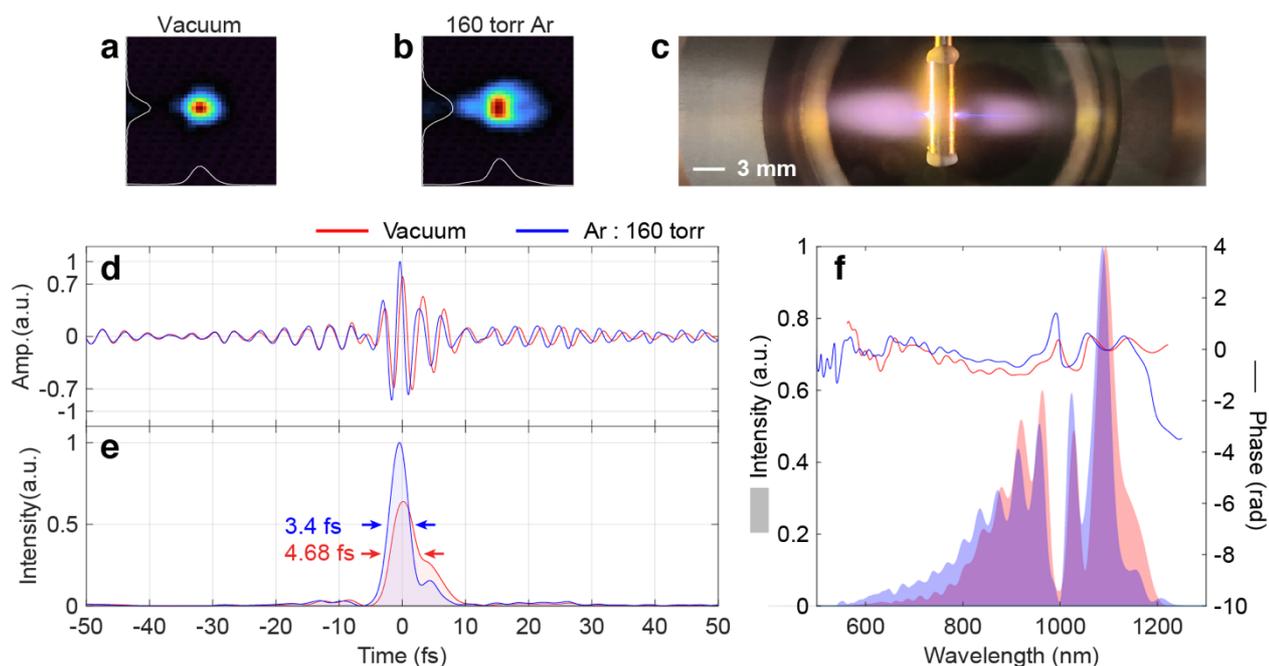

**Extended Data Fig. 4 | Pulse profiles after short gas cell (SGC).** (**a**) and (**b**) show the mode profile of the IR beam at the exit of the gas cell, using relay imaging, with the Ar pressure in the SGC at 0 torr and 160 torr, respectively. 160 torr is the optimal pressure for generating the brightest HHG. (**c**) Side view of the Ar-filled SGC. (**d**) The waveform of the few-cycle pulse propagating through a SGC in vacuum (red lines) versus when filled with 160-torr Ar (blue lines). The curves are relatively normalized to the time point t = -50 fs. (**e**) and (**f**) show the corresponding intensity profile and spectrum, respectively. Due to the ionization-induced self-compression effect, the pulse shortens from 4.68 fs to 3.4 fs. A noteworthy comparison can be made between the SGC and SIGC. In the SGC, the pulse undergoes self-compression, similar to that observed in the SIGC. However, while the IR beam profile in vacuum remains nearly ideal, the introduction of gas leads to significant beam expansion and a tendency toward splitting. This contrasts with the self-cleaning and self-guiding effects observed in the SIGC, as shown in Fig. 2. These differences underscore the limitations of the SGC in effectively spatiotemporally shaping the IR beam, resulting in a lower HHG yield compared to the SIGC (see Fig. 1f).

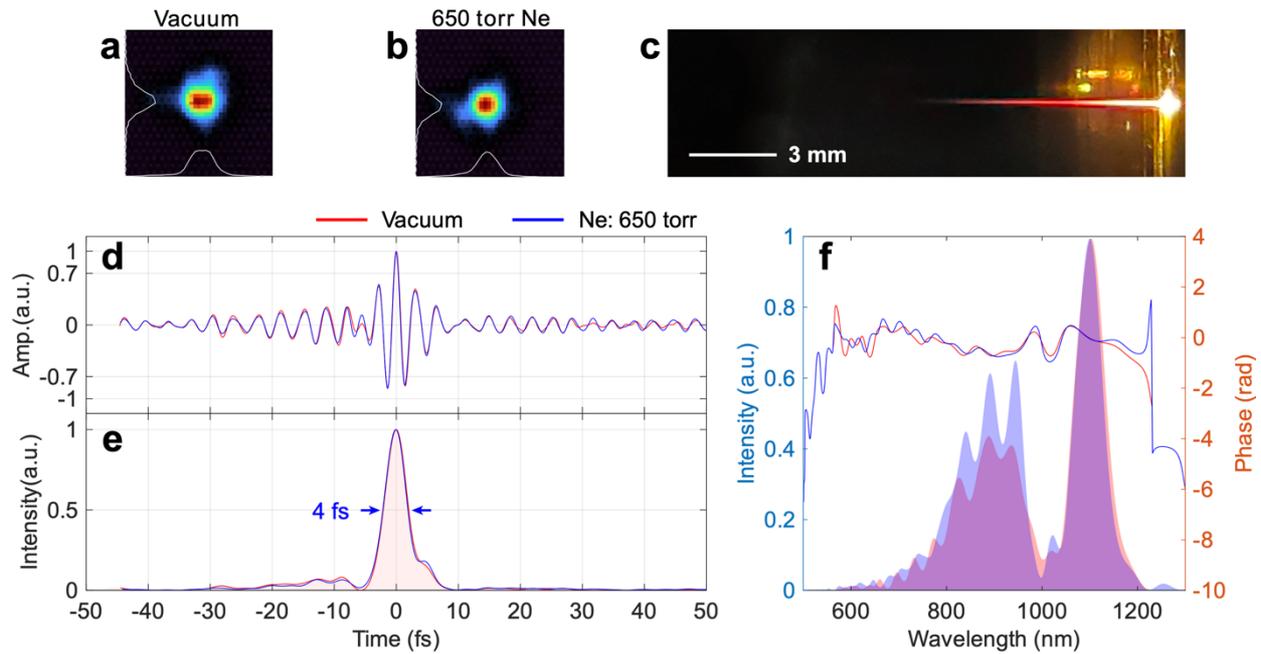

**Extended Data Fig. 5 | Self-cleaning effect in SIGC filled with Ne.** (**a**) and (**b**) show the mode profile of the IR beam at the exit of the gas cell, using relay imaging, with the Ne pressure in the SIGC at 0 torr and 650 torr, respectively. When Ne is present, there is a reduction in beam size and an improvement in beam quality, demonstrating the filamentation-induced beam guidance. (**c**) Side view of a filament through a 650-torr Ne-filled SIGC. The length of the plasma is about 4 mm. (**d**) The waveform of the few-cycle pulse propagating through a SIGC in vacuum (red lines) versus when filled with 650-torr Ne (blue lines), which is the optimized pressure for producing the brightest high-harmonic supercontinuum. The curves are relatively normalized to the time point t = -40 fs. (**e**) and (**f**) show the corresponding intensity profile and spectrum, respectively. The observed blue shift does not result in significant compression of the pulse duration, which remains around 4 fs with and without Ne.

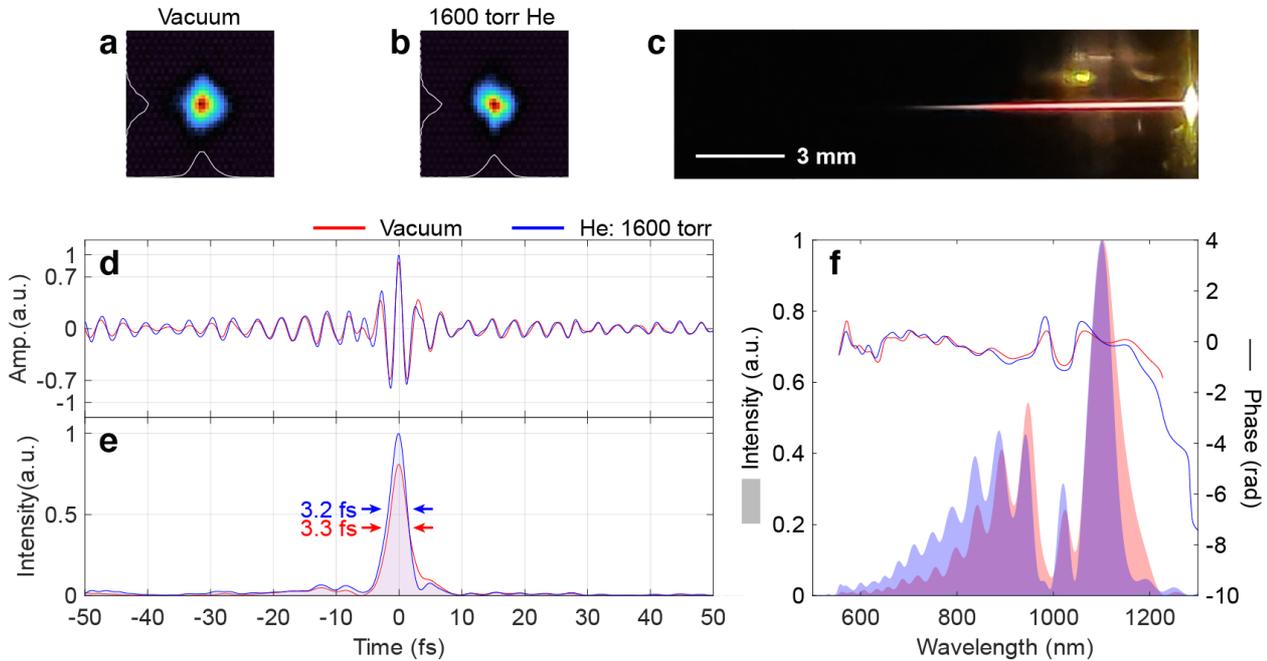

**Extended Data Fig. 6 | Pulse self-compression and self-cleaning effect in SIGC filled with He.** (**a**) and (**b**) show the mode profile of the IR beam at the exit of the gas cell, using relay imaging, with the He pressure in the SIGC at 0 torr and 1600 torr, respectively. 1600 torr is the optimized pressure for producing the brightest high-harmonic supercontinuum. (**c**) Side view of a filament through a 1600-torr He-filled SIGC. The length of the plasma is about 8 mm. (**d**) The waveform of the few-cycle pulse propagating through a SIGC without (red lines) and with He (blue lines). The curves are relatively normalized to the time point t = -20 fs. (**e**) and (**f**) show the corresponding intensity profile and spectrum, respectively. This blue-shift results in pulse self-compression from 3.3 fs to 3.2 fs.

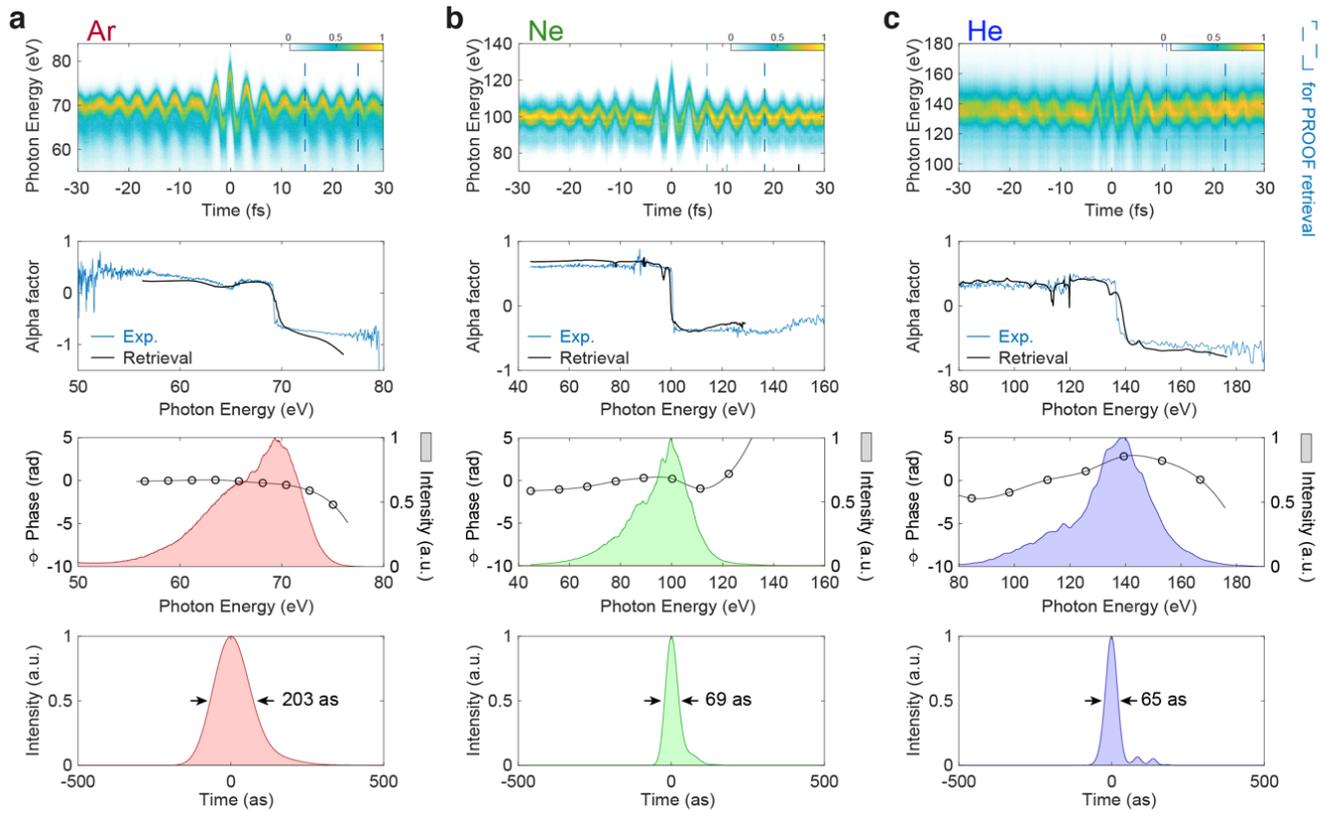

**Extended Data Fig. 7 | Retrieval of the resulting attosecond pulse with PROOF.** Top panel: The experimental spectrograms obtained by self-compressed IR and the resulting IAPs from the filaments of (**a**) Ar, (**b**) Ne, and (**c**) He. The blue dashed lines indicate the regions where the dressed field is monochromatic and weak, suitable for PROOF retrieval. Second column panel: Comparison of the phase angle $\alpha^{48}$ extracted from the experiment (blue lines) to that from the best retrieval (black lines). Third column panel: The PROOF-retrieved spectrum (filled area) and spectral phase (black line with circles). Bottom panel: Retrieved temporal intensity profile showing the pulse duration in FWHM.

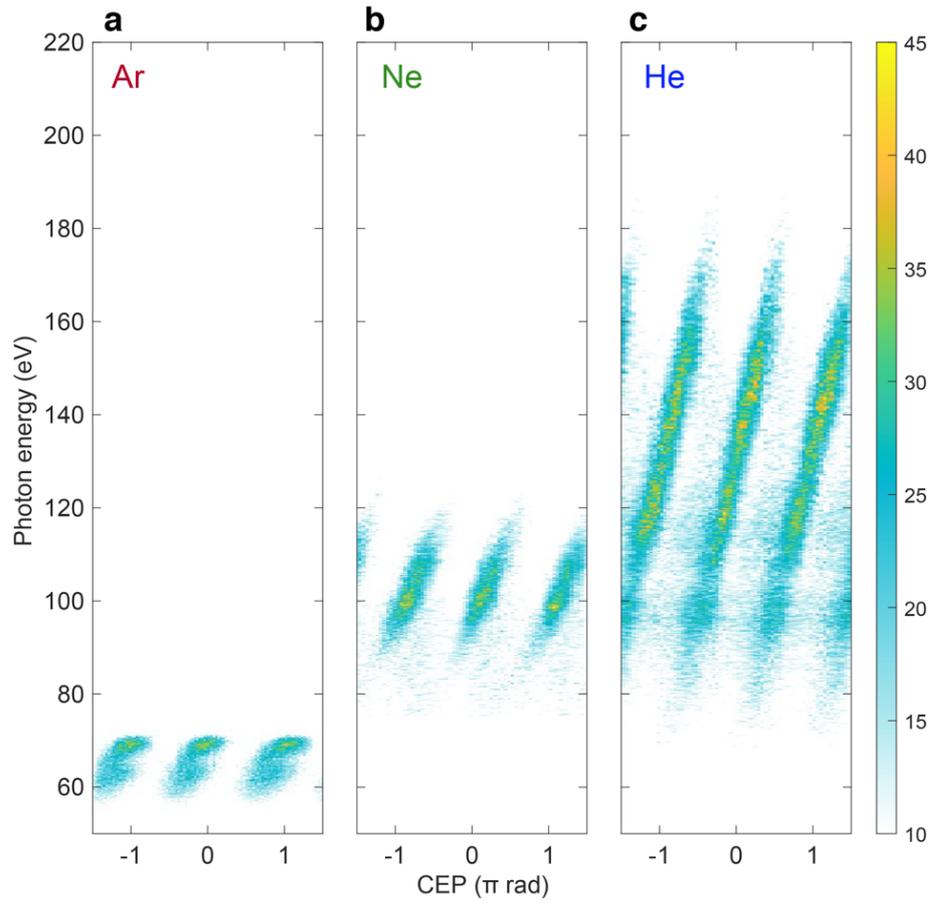

**Extended Data Fig. 8 | CEP-dependent photoelectron spectra.** CEP scan of photoelectron, i.e. EUV, continua generated in **a**, Ar (**b**, Ne, and **c**, He) and transmitted through a 200-nm Al (Zr and Ag) filter. A controlled variation of the CEP translates to a modulation of the shape and cutoff of the EUV continuum.

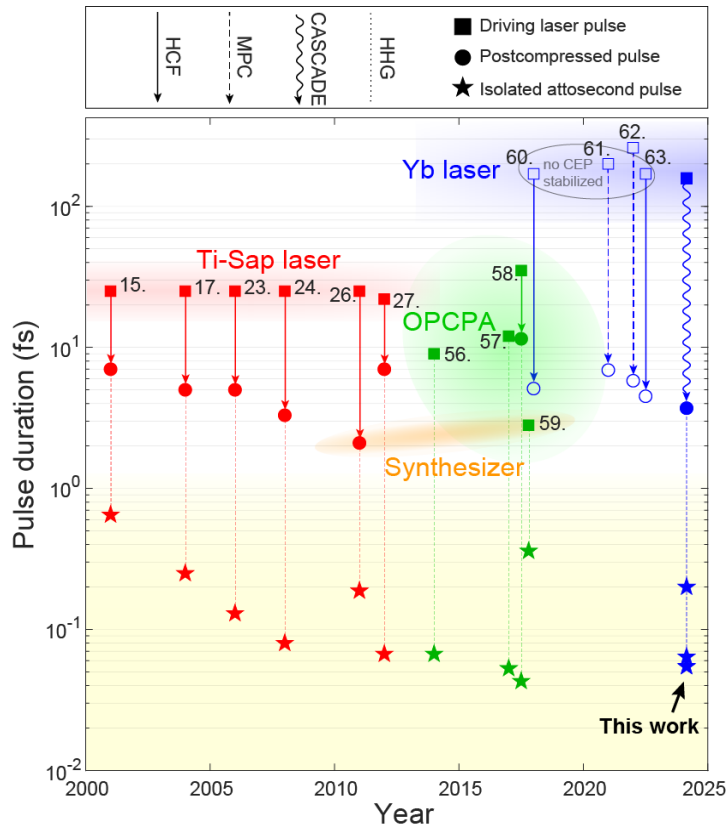

**Extended Data Fig. 9 | Laser sources and associated post-compression techniques demonstrated so far for generating IAPs.** There are four types of driving lasers currently in use: 1) CEP stabilized Ti:Sapphire lasers (red area), 2) CEP stabilized secondary sources, OPA (or OPCPA), driven by Ti:Sapphire or Yb lasers (green area), 3) multi-wavelength synthesizers developed for higher bandwidth (orange area), and 4) direct utilization of CEP stabilized Yb lasers (blue), which is presented in this work. The hollow core fiber (HCF) has been the major post-compression technology used for Ti:Sapphire and OPA (or OPCPA) lasers for IAPs. The multi-pass cell (MPC) is a recently developed post-compression method. However, since Yb lasers have a long initial pulse duration (>150 fs) and require a high post-compression factor (>40×) to achieve a one-cycle pulse for IAP generation, a new and simplified post-compression technique—CASCADE[37] combined with filamentation self-compression—is employed (see Extended Data Fig. 1).

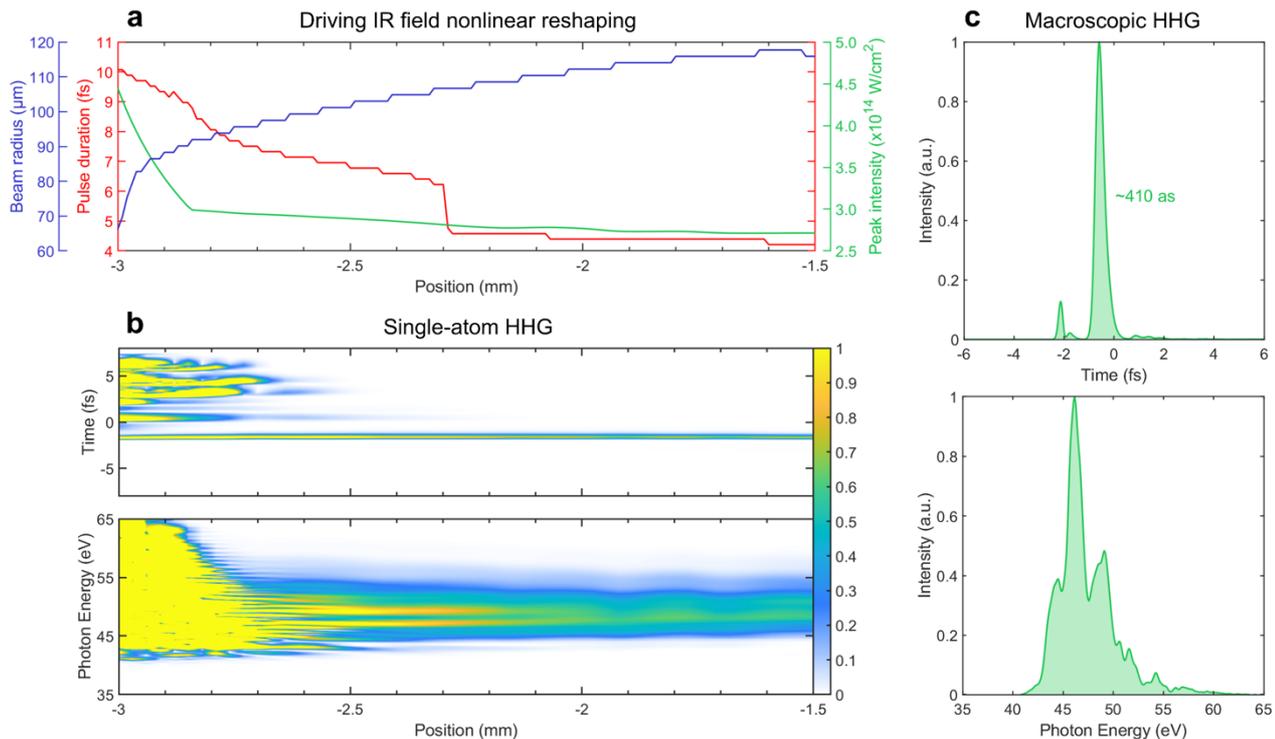

**Extended Data Fig. 10 | Numerical simulations of nonlinear driving field propagation and IAP generation in a SGC.** (**a**) Peak intensity (green) and size (blue) of the driving beam as it propagates along the final 1.5 mm of the 2.2-mm-long SGC filled with 120-torr Ar. The temporal FWHM of the driving pulse that propagates on-axis is plotted in red. For the simulation we used a beam with 335 μJ input energy with the same temporal profile as measured in the experiment and with a Gaussian spatial shape. With the SGC evacuated, the beam has a waist of 45 μm at the focus (position z = 0 in the horizontal axis), which is located 1.5 mm after the end of the SGC. (**b**) 3D-TDSE single-atom HHG calculations along the SGC propagation axis reflecting the attosecond temporal emission (top panel) and the corresponding harmonic spectra (bottom panel). (**c**) Attosecond pulse (top) and its EUV spectrum (bottom) obtained from the full macroscopic HHG simulations after propagation to a far-field detector. These macroscopic calculations predict the emission of a prominent isolated attosecond burst with a FWHM duration of ~410 as. Compared to the SIGC, pulse shaping in the SGC primarily occurs within the first ~1 mm of propagation, where the pulse peak intensity is abruptly reduced due to strong plasma defocusing at the cell entrance.

**Reference**

1. Goulielmakis, E. *et al.* Real-time observation of valence electron motion. *Nature* **466,** 739–743 (2010).
2. Mashiko, H. *et al.* Multi-petahertz electron interference in Cr:Al. *Nature Communications* 1–6 (2018). doi:10.1038/s41467-018-03885-7
3. Siegrist, F. *et al.* Light-wave dynamic control of magnetism. *Nature* **571,** 240–244 (2019).
4. Lucchini, M. *et al.* Unravelling the intertwined atomic and bulk nature of localised excitons by attosecond spectroscopy. *Nature Communications* **12,** 1021 (2021).

5. Tengdin, P. *et al.* Direct light–induced spin transfer between different elements in a spintronic Heusler material via femtosecond laser excitation. *Science Advances* **6,** eaaz1100 (2020).
6. Hui, D. *et al.* Attosecond electron motion control in dielectric. *Nature Photon* **16,** 33–37 (2022).
7. Hofherr, M. *et al.* Ultrafast optically induced spin transfer in ferromagnetic alloys. *Science Advances* **6,** eaay8717 (2020).
8. Inzani, G. *et al.* Field-driven attosecond charge dynamics in germanium. *Nature Photon* **17,** 1059–1065 (2023).
9. Schafer, K. J., Yang, B., DiMauro, L. F. & Kulander, K. C. Above threshold ionization beyond the high harmonic cutoff. *Phys. Rev. Lett.* **70,** 1599–1602 (1993).
10. Corkum, P. B. Plasma Perspective on Strong-Field Multiphoton Ionization. *Phys. Rev. Lett.* **71,** 1994–1997 (1993).
11. Rundquist, A. *et al.* Phase-Matched Generation of Coherent Soft X-rays. *Science* **280,** 1412–1415 (1998).
12. Gaarde, M. B., Tate, J. L. & Schafer, K. J. Macroscopic aspects of attosecond pulse generation. *J. Phys. B: At. Mol. Opt. Phys.* **41,** 132001 (2008).
13. Popmintchev, T. *et al.* Phase matching of high harmonic generation in the soft and hard X-ray regions of the spectrum. *Proceedings of the National Academy of Sciences of the United States of America* **106,** 10516–10521 (2009).
14. Christov, I. P., Murnane, M. M. & Kapteyn, H. C. High-harmonic generation of attosecond pulses in the 'single-cycle' regime. *Phys. Rev. Lett.* (1997).
15. Hentschel, M. *et al.* Attosecond metrology. *Nature* **414,** 509–513 (2001).
16. Sola, I. J. *et al.* Controlling attosecond electron dynamics by phase-stabilized polarization gating. *Nature Physics* **2,** 319–322 (2006).
17. Sansone, G. *et al.* Isolated Single-Cycle Attosecond Pulses. *Science* **314,** 443–446 (2006).
18. Locher, R. *et al.* Versatile attosecond beamline in a two-foci configuration for simultaneous time-resolved measurements. *Rev. Sci. Instrum.* **85,** 013113 (2022).
19. Ferrari, F. *et al.* High-energy isolated attosecond pulses generated by above-saturation few-cycle fields. *Nature Photon* **4,** 875–879 (2010).
20. Chen, M.-C. *et al.* Generation of bright isolated attosecond soft X-ray pulses driven by multicycle midinfrared lasers. *Proceedings of the National Academy of Sciences of the United States of America* **111,** E2361–E2367 (2014).
21. Hernandez-Garcia, C. *et al.* Isolated broadband attosecond pulse generation with near- and mid-infrared driver pulses via time-gated phase matching. *Opt. Express* **25,** 11855–11866 (2017).
22. Baltuska, A. *et al.* Attosecond control of electronic processes by intense light fields. *Nature* **421,** 611–615 (2003).
23. Goulielmakis, E. *et al.* Single-cycle nonlinear optics. *Science* **320,** 1614–1617 (2008).
24. Kaneshima, K. *et al.* Carrier-envelope phase-dependent high harmonic generation in the water window using few-cycle infrared pulses. *Nature Communications* **5,** 1–6 (2014).
25. Teichmann, S. M., Silva, F., Cousin, S. L., Hemmer, M. & Biegert, J. 0.5-keV Soft X-ray attosecond continua. *Nature Communications* **7,** 1–6 (2016).
26. Li, J. *et al.* 53-attosecond X-ray pulses reach the carbon K-edge. *Nature Communications* **2017,** 1–4 (2017).
27. Gaumnitz, T. *et al.* Streaking of 43-attosecond soft-X-ray pulses generated by a passively CEP-stable mid-infrared driver. *Opt. Express* **25,** 27506 (2017).
28. Sutherland, J. R. *et al.* High harmonic generation in a semi-infinite gas cell. *Opt. Express* **12,** 4430–4436 (2004).
29. Steingrube, D. S. *et al.* Generation of high-order harmonics with ultra-short pulses from

filamentation. *Opt. Express* **17,** 16177–16182 (2009).
30. Vismarra, F. *et al.* Isolated attosecond pulse generation in a semi-infinite gas cell driven by time-gated phase matching. *Light Sci Appl* **13,** 197 (2024).
31. Chin, S. L. *Femtosecond laser filamentation*. (Springer, 2010).
32. Couairon, A., Chakraborty, H. S. & Gaarde, M. B. From single-cycle self-compressed filaments to isolated attosecond pulses in noble gases. *Phys. Rev. A* **77,** 053814 (2008).
33. Chakraborty, H. S., Gaarde, M. B. & Couairon, A. Single attosecond pulses from high harmonics driven by self-compressed filaments. *Opt. Lett.* **31,** 3662–3664 (2006).
34. Shim, B., Nagar, G. C., Wu, Y. & Chang, Z. Generation of high-order harmonics and attosecond pulses in the water window via nonlinear propagation of a few-cycle laser pulse. *Opt. Express* **31,** 32488–32503 (2023).
35. Odhner, J. & Levis, R. J. Direct phase and amplitude characterization of femtosecond laser pulses undergoing filamentation in air. *Opt. Lett.* **37,** 1775–1777 (2012).
36. Kretschmar, M. *et al.* Direct observation of pulse dynamics and self-compression along a femtosecond filament. *Opt. Express* **22,** 22905–22916 (2014).
37. Tsai, M.-S. *et al.* Nonlinear compression toward high-energy single-cycle pulses by cascaded focus and compression. *Science Advances* **8,** eabo1945 (2022).
38. Itatani, J. *et al.* Attosecond Streak Camera. *Phys. Rev. Lett.* **88,** 173903 EP – (2002).
39. Weissenbilder, R. *et al.* How to optimize high-order harmonic generation in gases. *Nature Reviews Physics* **4,** 713–722 (2022).
40. Fibich, G. & Gaeta, A. L. Critical power for self-focusing in bulk media and in hollow waveguides. *Opt. Lett.* **25,** 335–337 (2000).
41. Loriot, V., Hertz, E., Faucher, O. & Lavorel, B. Measurement of high order Kerr refractive index of major air components. *Opt. Express* **17,** 13429–13434 (2009).
42. Weitenberg, J. *et al.* Multi-Pass-Cell-Based Nonlinear Pulse Compression to 115 Fs at 7.5 µJ Pulse Energy and 300 W Average Power. *Opt. Express* **25,** 20502–20510 (2017).
43. Holgado, W., Alonso, B., San Román, J. & Sola, I. J. Temporal and spectral structure of the infrared pulse during the high order harmonic generation. *Opt. Express* **22,** 10191–10201 (2014).
44. Park, S. B. *et al.* Direct sampling of a light wave in air. *Optica* **5,** 402–408 (2018).
45. Saito, N., Ishii, N., Kanai, T. & Itatani, J. All-optical characterization of the two-dimensional waveform and the Gouy phase of an infrared pulse based on plasma fluorescence of gas. *Opt. Express* **26,** 24591–24601 (2018).
46. Liu, Y. *et al.* All-optical sampling of few-cycle infrared pulses using tunneling in a solid. *Photon. Res.* **9,** 929–936 (2021).
47. Shelton, D. P. & Rice, J. E. Measurements and calculations of the hyperpolarizabilities of atoms and small molecules in the gas phase. *Chem. Rev.* **94,** 3–29 (1994).
48. Chini, M., Gilbertson, S., Khan, S. D. & Chang, Z. Characterizing ultrabroadband attosecond lasers. *Opt. Express* **18,** 13006–13016 (2010).
49. Hernández-García, C. *et al.* High-order harmonic propagation in gases within the discrete dipole approximation. *Phys. Rev. A* **82,** 033432 (2010).
50. Becker, A., Plaja, L., Moreno, P., Nurhuda, M. & Faisal, F. H. M. Total ionization rates and ion yields of atoms at nonperturbative laser intensities. *Phys. Rev. A* **64,** 023408 EP – (2001).
51. Durfee, C. G. *et al.* Phase matching of high-order harmonics in hollow waveguides. *Phys. Rev. Lett.* **83,** 2187–2190 (1999).
52. Popmintchev, T., Chen, M.-C., Arpin, P., Murnane, M. M. & Kapteyn, H. C. The attosecond nonlinear optics of bright coherent X-ray generation. *Nature Photon* **4,** 822–832 (2010).
53. Popmintchev, T. *et al.* Bright Coherent Ultrahigh Harmonics in the keV X-ray Regime from

Mid-Infrared Femtosecond Lasers. *Science* **336,** 1287–1291 (2012).
54. Kazamias, S. *et al.* High order harmonic generation optimization with an apertured laser beam. *Eur. Phys. J. D* **21,** 353–359 (2002).
55. Hernández-García, C., Sola, I. J. & Plaja, L. Signature of the transversal coherence length in high-order harmonic generation. *Phys. Rev. A* **88,** 043848 EP– (2013).
56. Kienberger, R. *et al.* Atomic transient recorder. *Nature* **427,** 817–821 (2004).
57. A, W. *et al.* Synthesized Light Transients. *Science* **334,** 195–200 (2011).
58. Zhao, K., Zhang, Q., Chini, M., Wu, Y. & Wang, X. Tailoring a 67 attosecond pulse through advantageous phase-mismatch. *Optics Letters* (2012).
59. Rossi, G. M. *et al.* Sub-cycle millijoule-level parametric waveform synthesizer for attosecond science. *Nature Photon* **14,** 629–635 (2020).
60. Jeong, Y.-G. *et al.* Direct compression of 170-fs 50-cycle pulses down to 1.5 cycles with 70% transmission. *Sci. Rep.* **8,** 11794 (2018).
61. Müller, M., Buldt, J., Stark, H., Grebing, C. & Limpert, J. Multipass cell for high-power few-cycle compression. *Opt. Lett.* **46,** 2678–2681 (2021).
62. Hädrich, S. *et al.* Carrier-envelope phase stable few-cycle laser system delivering more than 100 W, 1 mJ, sub-2-cycle pulses. *Opt. Lett.* **47,** 1537–1540 (2022).
63. Jeong, Y.-G. *et al.* Guiding of Laser Pulses at the Theoretical Limit – 97% Throughput Hollow-Core Fibers – with subsequent compression to 1.3 cycles. in (ed. Légaré, F. T. T. B. J. B. T. A. D. N.) Tu2B.4 (Optica Publishing Group, 2022).